\documentclass[useAMS,usenatbib]{mn2e}

\usepackage{times}
\usepackage{amsmath}
\usepackage{graphicx}  
\usepackage{epstopdf} 

\def\aj{AJ}

\def\apj{ApJ}
\def\apjl{ApJL}
\def\apjs{ApJS}

\def\aap{A\&A}
\def\aapr{{A\&A~Rev.}}
\def\aaps{{A\&AS}}

\def\mnras{MNRAS}

\def\nat{Nature}

\title[Misaligned Angular Momentum in Cosmological Simulations]{Misaligned Angular Momentum in Hydrodynamic Cosmological Simulations: Warps, Outer Discs, and Thick Discs}

\author[R. Ro\v{s}kar et al.]{
Rok Ro\v{s}kar,$^{1}$\thanks{e-mail: \tt{roskar@astro.washington.edu}}
Victor P. Debattista,$^{2}\thanks{RCUK Fellow}$
Alyson M. Brooks,$^{3}$ 
Thomas R. Quinn,$^{1}$
\newauthor
Chris B. Brook$,^{2}$ 
Fabio Governato,$^{1}$ 
Julianne J. Dalcanton,$^{1}$
James Wadsley,$^{4}$
\\
$^{1}$Astronomy Department, University of Washington, Box 351580, Seattle, WA 98195, USA \\
$^{2}$Jeremiah Horrocks Institute, University of Central Lancashire, Preston, PR1 2HE, UK\\
$^{3}$California Institute of Technology, M/C 350-17, Pasadena, CA 91125, USA\\
$^{5}$Department of Physics and Astronomy, McMaster University, Hamilton, ON, L8S 4M1, Canada\\
}

\date{Submitted May 05 2010}
\begin{document}

\maketitle

\label{firstpage}

%
%

\begin{abstract}

We present a detailed analysis of a disc galaxy forming in a high-resolution fully cosmological simulation to investigate the nature of the outer regions of discs and their relevance for the disc formation process. Specifically, we focus on the phenomenon of misaligned disc components and find that the outer disc warp is a consequence of the misalignment between the inner disc and the surrounding hot gaseous halo. As the infalling cold gas sinks toward the centre of the galaxy, it is strongly torqued by the hot gas halo. By the time the fresh gas reaches the central disc-forming region its angular momentum is completely aligned with the spin of the hot gas halo. If the spin of the hot gas halo, in turn, is not aligned with that of the inner disc, a misaligned outer disc forms comprised of newly accreted material. The inner and outer components are misaligned with each other because they respond differently to infalling substructure and accretion. The warped disc feeds the main gas disc due to viscous angular momentum losses, but small amounts of star formation in the warp itself form a low-metallicity thick disc. We show that observations of resolved stellar populations in warped galaxies in the local universe could provide evidence for the presence of these processes and therefore indirectly reveal ongoing gas accretion and the existence of hot gas halos.

\end{abstract}


%
%

\section{Introduction}

A large fraction of the baryonic mass in a spiral galaxy resides in the disc component. The discs of massive spirals assemble primarily from the quiescent cooling of shock-heated gas after the last major merger (e.g. \citealt{Fall:1980, Brook:2004, Robertson:2006}), although depending on the mass of the galaxy, cold accretion also plays an important role in providing disc gas (e.g. \citealt{Keres:2005, Brooks:2009}). Lower angular momentum material is initially denser and closer to the centre of the galaxy, therefore cooling faster and accreting early. It may also be generated through angular momentum loss during the merging process \citep{DOnghia:2006}, which in the hierarchical structure growth paradigm is more vigorous during the early stages of mass assembly, and yields old spheroids in the centers of galaxies. Higher angular momentum material follows late, resulting in inside-out growth of galactic discs \citep{Larson:1976,white:1991}. Observationally, inside-out growth can be inferred from trends of decreasing mean stellar age as a function of radius \citep{de-Jong:1996, Bell:2000, MacArthur:2004, Williams:2009, Gogarten:2010}, or by measurement of specific star formation rates as a function of radius \citep{Munoz-Mateos:2007}. By extension, the outermost regions of discs are the sites of current disc assembly, and therefore invaluable laboratories for studies of spiral galaxy formation. 

The outer reaches of discs are inherently difficult to study observationally due to low surface brightnesses several disc scale-lengths from the galactic centre. On the other hand, the high numerical resolution needed to study their properties hampers theoretical studies. However, it has long been acknowledged that these extreme regions of galaxies may hold unique clues to galactic disc formation. \citet{Freeman:1970} realized that the surface brightness profiles of some discs defy the usual (and simple!) characterization by a single exponential \citep{de-Vaucouleurs:1958} in their outer parts. Subsequent observations have revealed that downward-bending exponential profiles are very common, while pure exponential and upward-bending profiles constitute a smaller fraction \citep{van-der-kruit:1979, Kregel:2002,Pohlen:2002,Pohlen:2006}. 

The elusive outer discs present another puzzle - the extended HI is rarely aligned to the same plane as the inner disc. Misalignments between different components of spiral galaxies have been studied extensively over the past several decades \citep{Briggs:1990, Rubin:1994}. Phenomenologically, such misalignments range from warps, usually observed in HI \citep{Sancisi:1976, Garcia-Ruiz:2002, Verdes-Montenegro:2002} to structures whose angular momentum axes are orthogonal to that of the main disc, i.e. ``polar disc/ring'' galaxies (e.g. \citealt{Whitmore:1990}). In a modest sample of inclined galaxies, \citet{Garcia-Ruiz:2002} found that \textit{all} galaxies whose HI discs extend beyond the main stellar disc also have a warp. Warps have been shown to originate through torquing of the disc either by misaligned cosmic infall \citep{Ostriker:1989, Quinn:1992, Jiang:1999, Shen:2006}, misalignment between the angular momentum of the disc and the dark matter halo \citep{Debattista:1999}, or satellite perturbations \citep{Weinberg:2006}. The origin of polar ring structures has been the subject of much debate and there are many plausible mechanisms for their formation. Polar discs are often observed around S0 spirals and are postulated to form during an interaction (\citealt{Bekki:1998}) or  cosmological gas accretion \citep{Maccio:2006, Brook:2008}. 

Comparatively little work has been done on the stellar content coinciding with misaligned HI. \citet{Verdes-Montenegro:2002} showed that the warped outer layers also support low-level star formation in the case of NGC3642. Recent results from GALEX show faint UV emission in the warped discs of M83 \citep{Thilker:2005} and NGC5055 \citep{Thilker:2007, Sancisi:2008}, implying that a low level of star-formation also exists in those warps. In these cases, the UV emission is found well beyond the H$\alpha$ ``cutoff'' \citep{Kennicutt:1989}, resulting in an extended-UV (XUV) disc and at least circumstantially linking HI warps to this phenomenon.  Systematic studies of optically-identified warps agree with HI estimates of warp frequency and suggest that $\ga50\%$ of all galaxies are warped \citep{Sanchez-Saavedra:1990, Reshetnikov:1998}. 

The warps discussed above can only be observed in the local universe, but their ubiquity suggests that they are long-lived phenomena (unless they are all caused by recent satellite interactions). If large-scale HI warps harbor  sufficiently dense gas to allow for star formation, several generations of stars must have been born out of the plane of the main star forming disc. Furthermore, since discs grow with time, stars born out of the plane of the main disc must contribute to the stellar population not only in the tenuous outer regions, but throughout the disc, possibly contributing to a thick disc population. We address these intriguing possibilities below. 

We have previously addressed outer discs in the context of disc dynamics and stellar radial migration in (\citealt{Roskar:2008, Roskar:2008a}, R08 collectively hereafter). In this Paper, we focus on the importance of cosmological gas accretion and the intrinsic misalignments of different components for the outer parts of galactic discs. We study in detail the underlying cause for the misalignment between the central disc and the newly accreted gas, and find that it is the overall angular momentum of the entire gas reservoir which enables the misalignment. We therefore propose another explanation for the ubiquity of warps, and link this common feature of disc galaxies to their broader cosmological context.

The Paper is organized as follows: in $\S$~\ref{sec:methods} we briefly describe the simulation and the relevant details of the code; in $\S$~\ref{sec:misaligned_disk} we describe the morphological properties of the misaligned disc; in $\S$~\ref{sec:origin} we describe the physical origin of the misaligned disc in the simulation; in $\S$~\ref{sec:misalignedstars} we describe the properties of the stars that form in the misaligned disc; in $\S$~\ref{sec:thickdisk} we discuss the possible influence of long-lived warps on thick discs; in $\S$~\ref{sec:present_day_obs} we show that the processes described in the preceding sections may be detectable by resolved star studies of nearby systems; in $\S$~\ref{sec:dynamics} we discuss potential overlap of the warp signature imparted on the stellar populations to that of radial migration, and show that the disc is too hot to harbor sufficient non-axisymmetric structure needed for such secular evolution; we summarize our findings in $\S$~\ref{sec:discussion}.

%
%

\section{Methods}
\label{sec:methods}

The simulations presented in this Paper  were run using the $N$-body + SPH parallel tree code \textsc{gasoline} \citep{Wadsley:2004}, using Compton and radiative cooling, star formation and supernova feedback as described in \citet{Stinson:2006}, and a UV radiative background \citep{Haardt:1996}. The star formation recipe follows the prescription from \citet{Katz:1992}, forming stars according to 
\[
\frac{\text{d}\rho_{\star}}{\text{d}t}=c_{\star}\frac{\rho_{gas}}{t_{dyn}},
\]
where we use c$_{\star}=0.05$. $\rho_{gas}$ and $t_{dyn}$ are the local gas density and dynamical time evaluated using the SPH kernel. A gas particle is eligible to form stars if its density exceeds 0.1 amu/cm$^3$ and its temperature is below $1.5\times10^4$ K. The feedback recipe accounts for both SN II and SN Ia explosions, keeping track of iron and oxygen abundances. Supernova explosions impart energy upon particles enclosed within a blast radius based on \citet{Chevalier:1974}. To prevent the particles from immediately radiating away the energy, radiative cooling is turned off for a time corresponding to the end of the snow-plow phase of the SN remnant. 

We use two simulations in this paper. Our fiducial simulation, used to describe most of the physical processes in $\S$~\ref{sec:misaligned_disk}, $\S$~\ref{sec:origin} and $\S$~\ref{sec:misalignedstars} is the MW1 run discussed already in \citet{Brooks:2007, Pontzen:2008, Read:2009, Brooks:2009}. A lower-resolution version of the simulation was presented in \citet{Governato:2007, Governato:2008}. The reader is referred to those works for details on the initial conditions, here we just briefly outline the salient points. The simulation was run with WMAP year 1 cosmological parameters for the $\Lambda$-dominated, flat cosmology: $\Omega_0 = 0.3, \Lambda = 0.7, \Omega_b = 0.039, h = 0.7$. At z=0, the total mass within $R_{vir} = 270$ kpc is $M_{vir} = 1.1\times10^{12}$ M$_{\odot}$, resolved by a total of $1.26\times10^6, 2.85\times10^6$ and $6.7\times10^5$ dark matter, star, and gas particles respectively. The particle masses are $1.4\times10^5$ M$_{\odot}$ and $7.6\times10^5 M_{\odot}$  for the gas and dark matter respectively. Stars form with a mass of $4.8\times10^4 M_{\odot}$. The softening lengths of all particles are 0.3 kpc. Low-temperature cooling processes were not included in this simulation, limiting the minimum gas temperature to $\sim10^4$ K. 

Our fiducial simulation described above is an example of a system that was warped in the past but is entirely co-planar at z = 0. Therefore, we focus in $\S$~\ref{sec:present_day_obs} on another simulation (h603), which preserves its large-scale warp to the present day. The simulation was also run with GASOLINE, using WMAP year 3 cosmological parameters ($\Omega_0 = 0.24, \Lambda = 0.76, h = 0.73$). Its total mass within $R_{vir} = 186$ kpc is $M_{vir} = 3.7\times10^{11}$ M$_{\odot}$. The initial conditions were sampled on a $2304^3$ grid, resulting in $5.4\times10^5$ gas, $2.2\times10^6$ star, and $1.0\times10^6$ dark matter particles within the central galaxy's virial radius at z = 0. Initial particle masses for the gas and dark matter particles (in the central region) are $6.3\times10^4$ M$_{\odot}$ and $3.0\times10^5$ M$_{\odot}$ respectively. Initial stellar masses are $2.1\times10^4$ M$_{\odot}$. Softening lengths are 0.23 kpc for all particles in the inner region. This simulation uses a combination of low-temperature metal cooling \citep{Mashchenko:2006} and self-shielding sub-grid prescriptions, both of which are mostly relevant for evolution at high redshift, which is not the focus of this Paper. The star formation recipe and its parameters are identical to the fiducial MW1 run. All sections that follow with the exceptions of $\S$~\ref{sec:present_day_obs} and $\S$~\ref{sec:discussion} refer only to the fiducial run, MW1. 

%
%
\section{Morphology of the Misaligned Disc}
\label{sec:misaligned_disk}

%
%

%
%

\begin{figure*}
\centering
\includegraphics[width=6.5in]{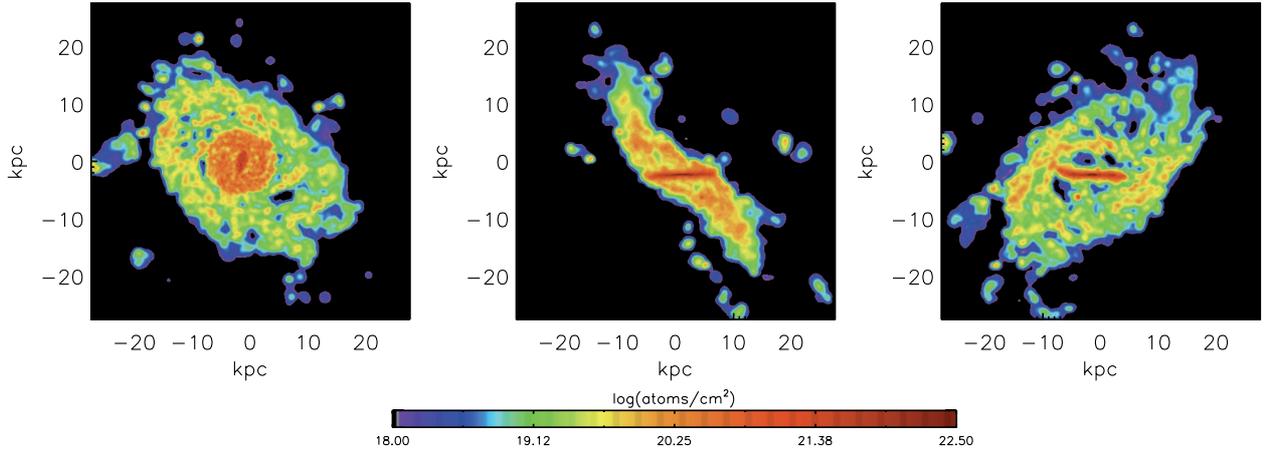}
\caption{HI maps of the disc at $z\sim0.5$ in a few different orientations. Colour indicates logarithmic HI column 
density, distance scale is in 
comoving 
kpc. The main disc is seen as the high-density central region in the inner $\sim10$ kpc.}
\label{fig:HI_map}
\end{figure*}

A misaligned disc is evident in our fiducial simulation almost immediately upon the completion of the last major merger at $z\sim2$, and persists for $\sim7$~Gyr.  In Figure~\ref{fig:HI_map} we show HI column density maps showing the central disc and the extended warp at z $\sim 0.5$. To measure the tilt, we rotate the simulation so that the angular momentum axis of the gas in the inner 3 comoving kpc is the $z$-axis. We then measure the angular momentum of gas in concentric annuli and take the tilt to be the angle between the angular momentum axis of the annulus and the $z$-axis. Note that this definition of warping differs from some definitions used in the literature to characterize observed warps (e.g. \citealt{Garcia-Ruiz:2002}), but we are not seeking exact model matches to the data; the measurement of tilt is purely for the determination of misalignment in the simulation and defining the tilt in terms of the angular momentum axis is a natural choice. In the analysis that follows, we define the warp to begin at a radius where the tilt angle exceeds $10^{\circ}$. 

\begin{figure}
\centering
\includegraphics[width=3in]{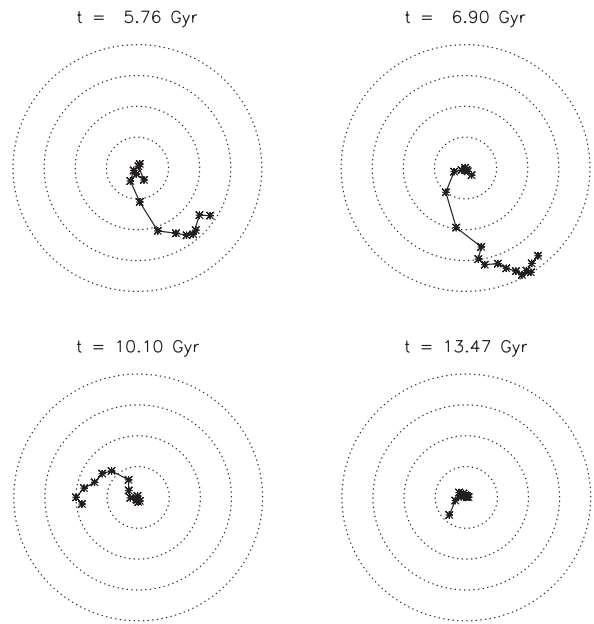}
\caption{Briggs figures at several different times. Each point represents a radial bin. There are a total of 15 bins spaced equally between 0 and 10 kpc. The radial coordinate in the polar representation is the angle of the angular momentum axis of the bin from the angular momentum axis of the main star-forming disc. The angular coordinate is the line of nodes (LON). The dotted circles are plotted in intervals of $10^\circ$. Direction of disc rotation is counter-clockwise.}
\label{fig:briggs}
\end{figure}

In Figure~\ref{fig:briggs} we show the orientation of the angular momentum vectors of concentric radial bins with respect to the main star-forming disk using Briggs figures. Briggs figures were introduced by \citet{Briggs:1990} to easily visualize the radial evolution of a warp by representing each radial bin as a point whose radial coordinate is the tilt (relative to the angular momentum of the central region) and the polar coordinate is the angle of the line-of-nodes (LON) from the $x$-axis. The LON is the imaginary line where the warp intersects the $xy$-plane\footnote{The polar angle shown is actually the angle between the angular momentum vector of the bin and the x-axis, which is 90$^\circ$ from the LON. However, since we are primarily interested in the change from one radius to the next rather than absolute angles, this difference in definition is irrelevant.}. From this representation, it is clear that the misalignment can in fact be considered a warp, given that successive rings twist and tilt with increasing radius. If, for example, the outer misaligned disk was an independent flat structure, we would see two clusters of concentric points in Fig.~\ref{fig:briggs}. The warp is already quite pronounced at $\sim$ 5.7 Gyr, reaching a tilt angle of $\sim 40^{\circ}$ by 7 Gyr. The warp gradually subsides and almost completely disappears by the present-day (we discuss the reasons for the disappearance of the warp in $\S$~\ref{sec:onsetend}). When a strong warp is present, it always traces out a leading spiral in the tilt-LON representation, consistent with the properties of observed warps \citep{Briggs:1990}. 

%
%

Our warp definition necessarily fails to identify a weak warp (such as in the bottom right panel of Figure~
\ref{fig:briggs}), but we find it impossible to reliably separate the warped layers from the unwarped disc when the 
tilting is $\la5^{\circ}$. A perfect decomposition is impossible because the transition between the 
two components is smooth, and the misaligned disc in fact feeds the main star-forming disc. However, as we 
show below, the gross properties of the large-scale warp that persists for a substantial portion of the simulation 
can be related to large misalignments between different galactic components, the understanding of which is not 
predicated upon a precise distinction of warped and unwarped parts of the disc.

\section{Origin of the Misaligned Disc}
\label{sec:origin}

Understanding the physical cause of the misalignment is crucial because of its potential to link an observational 
property of a galaxy to its broader, cosmologically relevant surroundings. Dynamical studies have successfully 
produced warps in simulations, by subjecting a flat disc of concentric rings or particles to an external torque 
\citep{Ostriker:1989, Quinn:1992, Jiang:1999, Debattista:1999, Lopez-Corredoira:2002, Shen:2006, Sanchez-Salcedo:2006}. In the case of our simulation, the misalignment precedes the evolution in the main disc plane and is a consequence of the gas accretion.

Several recent theoretical studies have addressed the process of gas delivery to galaxies
(\citealt{Birnboim:2003,Keres:2005, Dekel:2006, Keres:2008, Ocvirk:2008, Dekel:2009, Dekel:2009a, 
Brooks:2009}, hereafter B09). If the halo is massive enough ($\ga 10^{11} M_{\odot}$)
to support a stable shock, some gas is shock-heated to the virial temperature upon entering 
the halo (hot accretion), while other gas penetrates the halo inside a filament and is never 
shock-heated (cold accretion). Because such cold flows could potentially deliver substantial 
amounts of specific angular momentum deep into the galactic potential well, misaligned 
components of disc galaxies have been invoked previously as manifestations of cold accretion 
\citep{Maccio:2006, Brook:2008, Sancisi:2008, Spavone:2010}.

While most of the above studies use a temperature criterion to separate the two modes of smooth accretion, B09 additionally used an entropy criterion on the basis that a strong shock will always induce an entropy jump, but not every high-temperature particle must necessarily have been shocked. In what follows we use the same particle groupings as B09, dividing the gas particles in the main galaxy halo into unshocked,  shocked and clumpy (accreted after being a part of another halo). These definitions are applied to particles more than 30 kpc 
from the central star-forming disc, so as to not confuse entropy jumps and temperature changes due to large-scale accretion shocks with those from supernova feedback within the star-forming disc. Below we explore whether the misaligned gas component in our simulation arises due to cold flows, dark matter halo torques, or whether it results from another mechanism.

%
%

\begin{figure}
\centering
\includegraphics[width=3in]{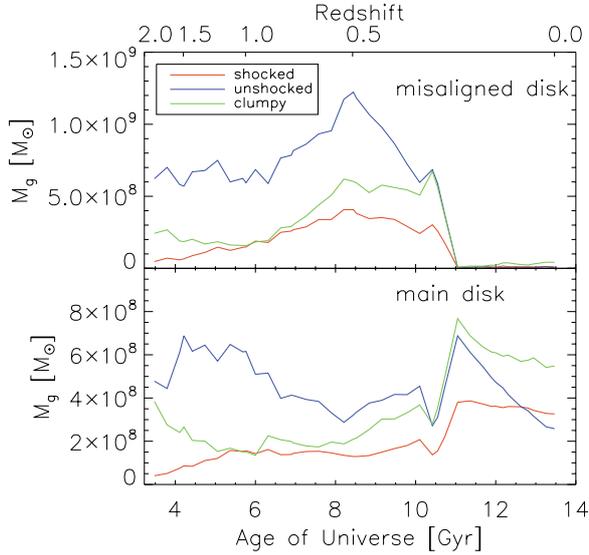}
\caption{Total gas mass in the misaligned disc and main star-forming disc, separated by mode of accretion. Red (shocked) corresponds to particles that enter the virial halo and shock at the virial radius, blue (unshocked) shows particles that enter the virial halo unshocked, and green (clumpy) shows particles 
that enter the virial radius after having belonged to another halo.
}
\label{fig:discmass}
\end{figure}

In Figure~\ref{fig:discmass} we show the gas mass found in each component (misaligned disc and main star-forming disc) as a function of time, broken down according to the mode of accretion, as discussed above. The total amount of gas found in the misaligned component is as much as a factor $\sim$~2 larger than that in the main star-forming disc. Importantly, though the misaligned disc is mostly composed of gas acquired through cold accretion ($\sim75$~per cent),  all modes of accretion contribute to its total mass. The misaligned gas therefore cannot only be a consequence of accretion of unshocked gas, but must also be due to a more general mechanism that affects all gas equally. In the following sections, we explore possible mechanisms for this behavior.

\subsection{Warps as Indicators of Misaligned Hot Gaseous Halos}
\label{sec:hothalomisalignment}

In this section, we propose that the outer disc warps are a result of the misalignment between the inner disc and the surrounding hot gas halo. We define ``hot'' and ``cold'' gas as all gas with $T > 10^5 \text{ K}$ and $T < 10^5 \text{ K}$ respectively. As the cold gas sinks toward the centre of the galaxy it is torqued by the massive hot gas halo long before it reaches the disc. Due to this torquing in the outer halo, the cold gas aligns with the spin of the hot gas halo by the time it reaches the central region. Because the hot gas halo is misaligned with the main star-forming disc, its misalignment manifests as a warp that is mainly composed of cold infalling gas. 

\begin{figure}
\centering
\includegraphics[width=3in]{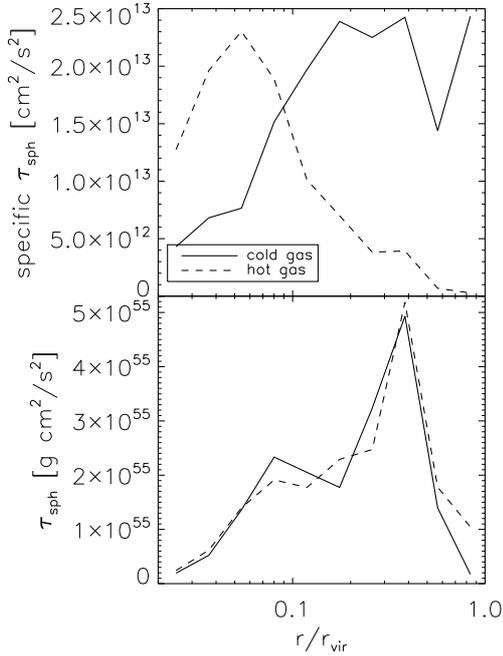}
\caption{Top and bottom panels show the specific and total hydrodynamic torques respectively on cold (solid lines) and hot gas (dashed lines) at z $\sim0.5$ (t $\sim8$ Gyr). The specific torques are much different because hot gas dominates by mass in the halo, however the absolute torque magnitudes are equal, showing that the hot gas torques the cold gas.}
\label{fig:sphtorques}
\end{figure}

In Figure~\ref{fig:sphtorques} we show explicitly that the hot gas halo is responsible for torquing the infalling cold gas. In the top panel, we show the specific torque resulting from hydrodynamic forces on cold and hot gas (solid and dashed lines respectively). The specific torque on the cold gas is much higher in the outer parts of the halo, because the hot gas dominates the cold gas by mass in the outer regions. However, the \emph{total} torque on each subset is almost exactly the same, as shown in the bottom panel. We have verified that the directions of the SPH torques on the two types of gas are antiparallel in the outer parts of the halo. Hence, the hot and cold gas exert a torque on each other, but the more massive hot gas halo dictates the sense of rotation for the infalling cold gas. Although we only show one timestep in Fig.~\ref{fig:sphtorques}, this behavior is evident at other timesteps we analyzed after $z=2$.

We have also investigated the evolution of hot gas halo gas present within the virial radius as a function of time. Although the angular momentum of the hot gas halo gas changes through accretion, the halo particles behave coherently. For gas that makes up the hot gas halo ($\text{T}>10^5$ and $0.3 < r/r_{vir} < 0.7$) at 5 Gyr, the angular momentum direction remains largely radially coherent at all times.

%
%

\begin{figure}
\centering
\includegraphics[width=3in]{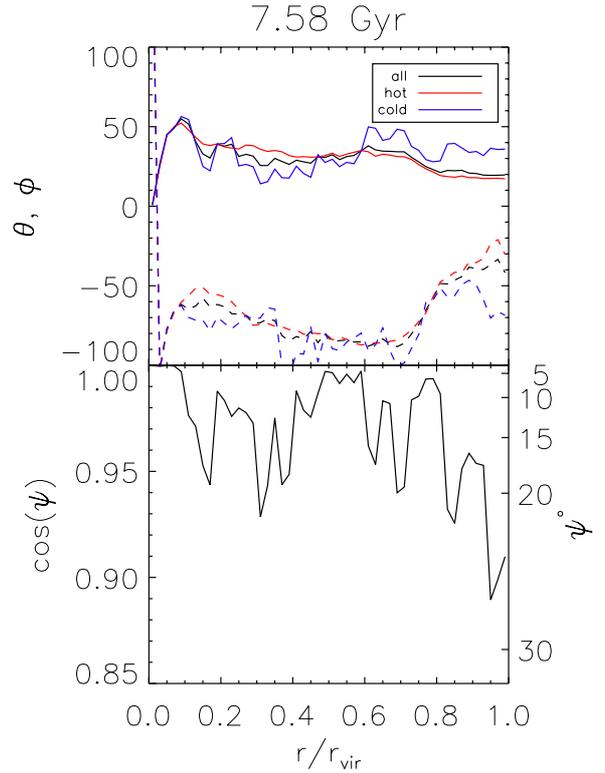}
\caption{{\bf Top:} Mean angles $\theta$ (solid lines) and $\phi$ (dashed lines) of gas angular momentum 
vector subdivided by gas temperature. Note that although the gas with lower temperature will sink quicker to 
the centre of the halo, the hot gas dominates by mass and drives the sense of rotation. The region of the 
misaligned (warped) disc is roughly between $0.05 \la r/r_{vir} \la 0.15$. 
{\bf Bottom:} Angle between the cold and hot gas angular momenta as a function of radius.
}
\label{fig:avg_angles}
\end{figure}

The torquing of cold gas results in very close alignment of angular momenta of the hot gas halo and infalling cold gas. In the top panel of Figure~\ref{fig:avg_angles} we show the radial profile of $\theta$ (solid lines) and $\phi$ (dashed lines), angles between the $z$ and $x$ axes of the main star-forming disc and the angular momentum vector of gas in each bin in a single snapshot. The gas is subdivided into cold and hot, again based on a temperature criterion regardless of whether it shocked or not, as we are interested in seeing whether cold gas preserves its trajectory through the halo\footnote{Selecting particles purely based on mode of accretion may confuse the analysis because it could include particles accreted as unshocked, but expelled from the disc through supernova feedback and therefore belonging to the hot gas halo. By limiting the selection according to present temperature, we make sure in part that the cold particles are ones currently falling toward the central disc for the first time. These may and do include particles that shocked upon accretion.}.
The bottom panel explicitly shows the alignment of the hot and cold gas defined by 
\begin{equation}
\psi = \cos^{-1}\left(\frac{\vec{J_1}\cdot \vec{J_2}}{|J_1||J_2|}\right).
\label{eq:dotproduct}
\end{equation}
The top panel shows much more variation in the profile of the cold gas, though on average the cold gas follows the hot gas, especially closer to the centre of the halo. The angle between the two components steadily decreases inwards (bottom panel) and they converge completely in the inner region where the disc fueling takes place.  

%
%

\begin{figure}
\centering
\includegraphics[width=3in]{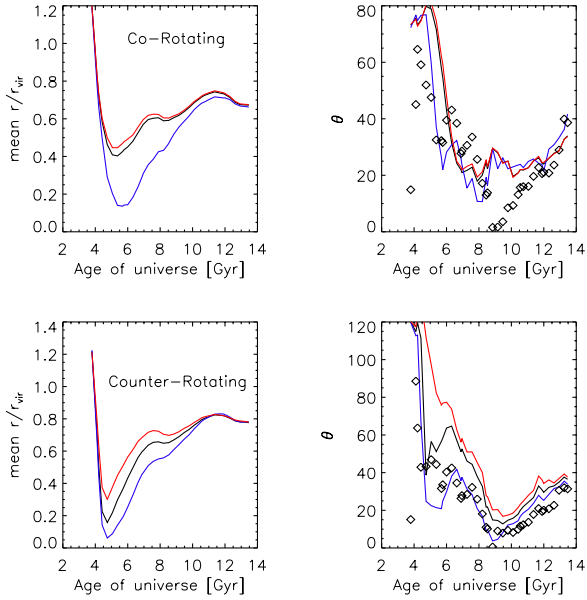}
\caption{Time evolution of particle properties for gas particles that enter the virial radius at the same time ($t \sim 3.8$ Gyr). Top and bottom panels show particles that upon entering the halo are co- and counter-rotating respectively. The particles were chosen from narrow regions of $\theta$ and $\phi$ distributions, where $\theta$ and $\phi$ specify the angular momentum vector of the individual particles with respect to the central disc (see text for more details). Black, blue, and red lines represent all, unshocked, and shocked particles respectively. Symbols in the right panels denote the rotation of the halo around the mean radius of the particles at a given time. Both sets of particles, co-rotating and counter-rotating, eventually align with the angular momentum of the hot gas halo. }
\label{fig:rvir_enter}
\end{figure}

In Figure~\ref{fig:rvir_enter},  we show the evolution of particles that enter the virial radius at $t\sim3.8$ Gyr, grouped  according to their angular momentum alignment upon entering the halo. We chose two very different groups of particles in order to investigate whether the incoming angular momentum is preserved as the gas sinks to the centre. We follow the mean properties of the particles in each timestep, though the interpretation is made more difficult by the fact that the particles do not necessarily remain grouped spatially. Nevertheless, this approach sheds some light on the gas evolution within the virial radius. 

Specifically, we select the particles according to their angular momentum vector with respect to the angular momentum of the central disc, yielding two subgroups related both spatially and by their direction of motion. Top and bottom panels show particles that were co- and counter-rotating as they entered the virial radius respectively. The range of angular momentum vector angles used for the selection are $118^{\circ} < \theta < 122^{\circ}$ and $-160^{\circ}<\phi<-140^{\circ}$ for counter-rotating particles and $70^{\circ} < \theta < 78^{\circ}$ and $50^{\circ}<\phi<70^{\circ}$ for the particles co-rotating with the disc. 

%
%

Even in the extreme case of unshocked counter-rotating gas entering the halo (bottom panels, blue lines) the sense of rotation with respect to the central disc is \emph{not} preserved, indicating that the infalling gas experiences substantial torques as it travels through the halo. To demonstrate that the hot gas halo is indeed the source of the torques, we compute the rotation of the hot gaseous halo within a shell of thickness 0.05~$r_{vir}$ about the mean radius of the selected particles, and show it in the right-hand panels with diamond symbols.\footnote{This is not ideal because, as we noted above, the particles we select initially do not necessarily remain tightly grouped. For example, the offset in $\theta$ between the shocked and unshocked particles at $\sim8$ Gyr is due to them occupying slightly different parts of the halo. Nevertheless, this comparison is useful for relating the infalling particles to the halo in a general sense.} The infalling cold particles in general follow the halo rotation quite closely, in agreement with the expectations based on Fig.~\ref{fig:sphtorques}. 

Figure~\ref{fig:rvir_enter} also shows that the gas that entered the halo in the hot phase sinks to the centre more slowly, and a smaller fraction makes it to the star forming region (left-hand panels of Fig.~\ref{fig:rvir_enter}). About 50 per cent of the unshocked gas and about 30 per cent of the shocked gas gets consumed by star formation before being heated by supernova feedback and ejected back into the hot gas halo. Once the gas is ejected from the disc, it becomes a part of the hot gas halo at temperatures $>10^6$ K, which is evident by the rapid increase in mean $r$. Also note that the large discrepancy in the alignment at 8-9 Gyr is due to an infalling satellite which complicates a robust determination of angular momentum angles at those times.  

We note that a qualitatively similar phenomenon was pointed out by \citet{Katz:1991}, who also found that the outer parts of the gas disc were warped with respect to the centre, but aligned with the angular momentum vector of the initial top-hat distribution. They attributed the offset to the differences in the details of angular momentum transfer between the gas blobs and the dark matter halo. The ``angular momentum catastrophe'' in today's state-of-the-art simulations has been alleviated by increased resolution and more sophisticated modeling, but in the fully-cosmological context, the added complications of dark matter halo triaxiality and merging facilitate the perpetuation of misalignment.

\subsection{Misalignment due to Dark Matter Halo Torques}
\label{sec:dmtorques}

Though we have shown above that the hot gaseous halo torques the infalling cold gas, the torquing by the gravitationally dominant DM halo could dictate the overall angular momentum properties of the gas and ultimately, the misaligned disc. Two main possibilities exist: a) the DM efficiently torques the gas particles, influencing their angular momentum alignment in the halo and b) the DM halo torques the flat disc into a warp. We address each of these in turn below.

Investigating the first of these two possibilities, we first measure the shape of the dark matter halo in ellipsoidal shells using an iterative procedure. On the first iteration, the mass moment tensor, defined by
\begin{equation}
\mathbf{M}_{j,k} = \frac{1}{M}\sum_{i} m_i r_j r_k
\end{equation}
in a given spherical shell is diagonalized to obtain the principal axis vectors and the 
associated eigenvalues. Note that this is not the moment of inertia tensor, although 
in the literature this nomenclature is often used (see \citealt{Bett:2007}). On subsequent 
iterations, we redefine the ellipsoidal shell according to the eigenvectors by

\begin{equation}
r_{ell} = \sqrt{x^2  + \left(\frac{y}{b/a}\right)^2 + \left(\frac{z}{c/a}\right)^2},
\end{equation}
after rotating the particle coordinates into the frame where the mass distribution matrix 
is diagonal. The axis lengths $a$, $b$, and $c$ are given by the square roots of the 
eigenvalues. Once the shell is redefined, the mass distribution matrix is reconstructed and the procedure 
repeated. We find that in most cases the deviations in axis ratios are less than 1 per cent after $\sim20$ iterations. 
This procedure is analogous to that used in \citet{Katz:1991a}, though it uses 
differential instead of cumulative shells following \citet{Debattista:2008}. 

For the majority of the simulation from $z\sim2$ to the present, the axis ratios of the DM halo are $b/a\sim0.8-0.9$ and $c/a\sim0.5-0.7$ beyond $r/r_{vir} \sim 0.2$. We find that in the innermost region the angular momentum of the gas is well-aligned with the minor axis of the DM halo (also reported by \citealt{Bailin:2005a}), but we do not find any obvious coupling between the gas angular momentum vector and the dark matter halo's minor axis during periods when the warp is present. 

\begin{figure}
\centering
\includegraphics[width=3in]{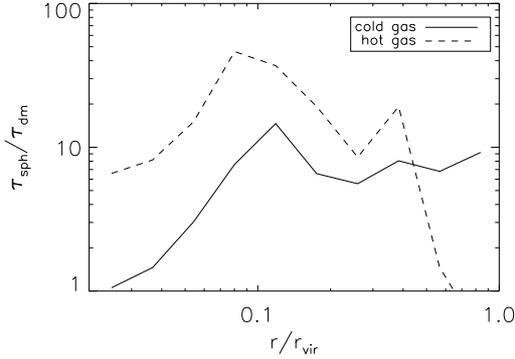}
\caption{The ratio of SPH torque to DM torque on the cold and hot gas in logarithmically-spaced radial bins at z$\sim$0.5. Except for the innermost bins in the region of the central disc, the SPH torque dominates over that of the dark matter.} 
\label{fig:torques}
\end{figure}

To check explicitly the importance of DM torques on the trajectories of gas particles through the halo, we calculate gravitational and hydrodynamic (SPH) torques on the particles directly.  Figure~\ref{fig:torques} shows  the comparison of torques from DM and hydrodynamics on the cold and hot gas. Recall that the total SPH torques were equal on these two subsets of particles (Fig.~\ref{fig:sphtorques}). Fig.~\ref{fig:torques} shows explicitly that the DM torques are insignificant in shaping the trajectory of gas particles through the halo. Differences between the hot and cold gas arise because the cold gas is in general patchy and irregularly spaced in each radial shell, therefore experiencing a larger net gravitational torque in each shell from the dark matter halo than the more homogeneous and spherical hot halo gas. 

%
%

This result seems counter intuitive, given that the dark matter dominates the potential of the galaxy, but a simple experiment confirms that such behavior is expected. To simplify the problem, we imagine a triaxial halo of uniform density at the origin, with principal axis lengths $a$, $b$, and $c$ (aligned along $x$, $y$, and $z$ axes respectively), where $a > b > c$. We assign a mass $M_{halo} = 3\times10^{11} M_{\odot}$ to the halo and set the longest axis to be $a = 0.5 r_{vir}$. The other two axis lengths are set according to axis ratios measured in the simulation with $b/a = 0.8$ and $c/a = 0.6$. We then place rings of test masses at a distance $r_{c}$ to represent the clouds of infalling material. The two rings are inclined with respect to the $xy$-plane by $45^\circ$, rotated around the $x$ and $y$ axes respectively. We may then calculate the gravitational torques $\tau_g$ on the test masses in the rings using equations for potentials of homogeneous spheroids given in Table 2.1 and 2.2 of \citet{Binney:2008}. 

Similarly, we can make a simple estimate for the expected ram pressure force on a cloud of gas moving through the rotating hot gas halo. The torque due to the ram pressure 
$P_{ram} \approx \rho_{h}v_{rel}^2$ is given by
\begin{equation}
\vec{\tau}_{ram} = \vec{r_c} \times (P_{ram}A_{c}\hat{v}). 
\label{eq:ram_torque}
\end{equation}
We assume in this estimate that the clouds in our simulation are analogous to compact high velocity clouds (CHVCs) observed in the halo of the Milky Way and other nearby systems \citep[e.g.][]{Westmeier:2005, Putman:2002, Pisano:2007}. We assume that the clouds are moving with velocities 100-200 km s$^{-1}$ relative to the hot gas halo. Sizes of CHVCs are fairly unconstrained due to inherent difficulties in estimating their distances, though most constraints place HVCs within 150~kpc of the host galaxy centers \citep[see][]{Putman:2006}. To estimate the size of the clouds, we take the mean angular size CHVCs from \citet{Putman:2002}, $A_{sky} = 0.36 \text{ deg}^2$. Assuming that the clouds are at 150 kpc, we obtain a radius of $\sim1$~kpc. We consider this to be an upper limit and consider radii in the range of 0.05 - 1.0~kpc. Using equation~\ref{eq:ram_torque} and placing the clouds anywhere between $0.6 r_{vir} < r_{c} < r_{vir}$, we find that $\tau_g/\tau_{ram} \la 30$~per cent for reasonable choices of halo and cloud parameters. The behavior seen in the simulation, where gas trajectories are determined largely by hydrodynamic torques, is therefore expected for cold gas encountering a hot gas halo in the outer parts of the galaxy.

We now try to determine whether the warp could be caused by the torques of the \emph{outer} halo on a flat disc, as was recently demonstrated as a plausible warp formation mechanism by \citet{Dubinski:2009}. We calculated the torque in the disc plane due to the DM halo particles beyond $r/r_{vir} > 0.2$ (i.e. beyond where the halo and the disc are aligned) in a few timesteps where the warp is clearly evident, including the output shown in Figure~\ref{fig:HI_map}. The timescale for changing the angular momentum $J$ of a ring at radius $r$ by $\Delta J \sim J$ is given by $t(r) = J(r)/\tau_{\bot}$, where $J$ is the angular momentum of the ring and $\tau_{\bot}$ is the component of the torque perpendicular to the ring's angular momentum vector. We find that this timescale is $> 10$ Gyr everywhere. Considering that we see a tilt of $\sim40^{\circ}$, $J$ only needs to change by  25 per cent, but the timescale for such change is still several Gyr in the region of the warp. If this were the dominant mechanism for the warping, we should first see a flat disc develop, whose outer regions would torque out of the plane slowly. This is not the case, as the warp is in place almost immediately after the last major merger. Therefore, the outer DM halo torque cannot produce the warp in this system. 

We conclude that warp formation in this simulation is not significantly affected by the dark matter torques because hydrodynamic torques dominate in the outer halo where the spin of the infalling gas is determined prior to its entry into the disc region. By the time the fresh gas reaches the central region, it is aligned with the hot gas  halo, but \emph{misaligned} with the central disc, giving rise to a warp. The warp  is therefore a manifestation of a misalignment between the spin axes of the central disc and the surrounding hot gaseous halo.    

\subsection{Onset and End of the Warp}
\label{sec:onsetend}

We have established in the previous section that the misaligned cold gas component in our simulation
is due to distinct spin axes of the inner disc and the surrounding hot gas halo. Here, we try to determine what causes the two components to be misaligned in the first place and why the warp eventually disappears. 

\begin{figure*}
\centering
\includegraphics[width=6.5in]{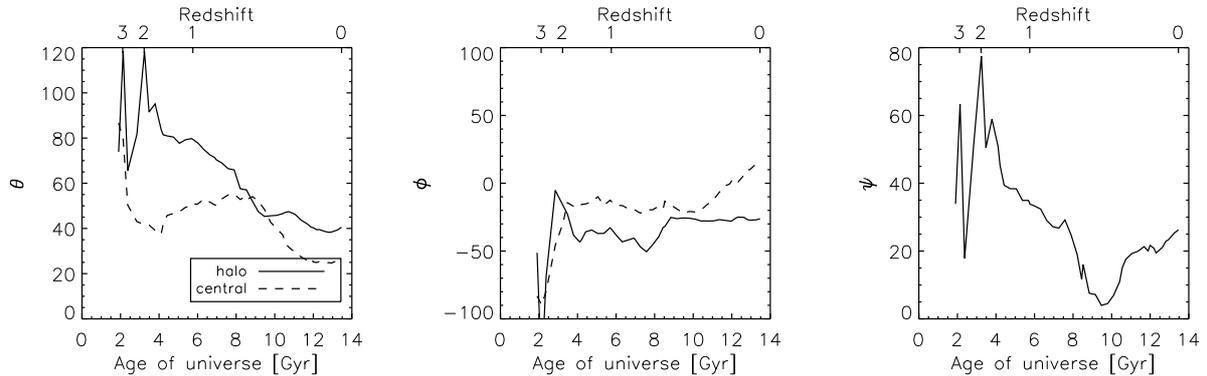}
\caption{Evolution of direction of angular momentum vectors of the main disc ($r < 3$~kpc) and the gas halo ($0.2 < r/r_{vir} < 0.5$). $\theta$ and $\phi$ are angles between the angular momentum vectors and the $z$ and $x$~-axes of the simulation box, respectively. $\psi$ is the angle between the angular momentum vectors of the two components. During the major merger at $z\sim2$, the evolution of the spin axes of the two components is decoupled, resulting in an overall misalignment between the central disc and the rest of the gas halo. }
\label{fig:lmm_align}
\end{figure*}

First, we measure the misalignment as a function of time. We define the inner disc as particles with $r < 3$ comoving kpc and we measure the halo properties in a shell between $0.2 < r/r_{vir} < 0.5$. We limit the shell to 0.5 $r/r_{vir}$ to minimize the variations  due to the newly infalling matter. In the left and centre panels of Figure~\ref{fig:lmm_align} we show the orientation of the gas halo (solid lines) and the main disc (dashed lines) with respect to the static simulation frame as a function of time. In the right panel we show $\psi$, the angle between the angular momentum vectors of the two components defined in equation~\ref{eq:dotproduct}. 

The two components enter the quiescent period of evolution misaligned by $\ga 50^{\circ}$, following the 
last major merger at z $\sim2$. The disc's orientation is essentially constant between 4-9 Gyr prior to the minor merger, which takes place at $\sim 9.5$~Gyr, but the halo orientation is slowly changing due to ongoing accretion. The minor merger provides an important clue as to why the inner disk and the outer gas halo are misaligned in the first place. The infalling satellite rapidly torques the disc, changing its orientation in the inertial frame and with respect to the hot gas halo. The gas halo, however, is unaffected by the merger because the infalling satellite is not massive enough to have a considerable effect on the halo structure or angular momentum content. The misalignment between the different components is therefore perpetrated by such interactions that affect the inner and outer regions differently. 

%
%

Figure~\ref{fig:discmass} shows that the warp ends at $\sim$11 Gyr, yet  Figure~\ref{fig:lmm_align} shows that even at $z=0$ the gas halo is misaligned with respect to the central star forming disc. If the warp is due to the misalignment, why does it disappear while the misalignment persists? Note that although the warp is long-lived, the gas never lingers there for more than 1-2 Gyr and it therefore must continually be supplied from the halo (see Figure~\ref{fig:rvir_enter}, left panels). Thus, a gaseous warp may disappear when it ceases to be resupplied from the halo. 

\begin{figure}
\centering
\includegraphics[width=3in]{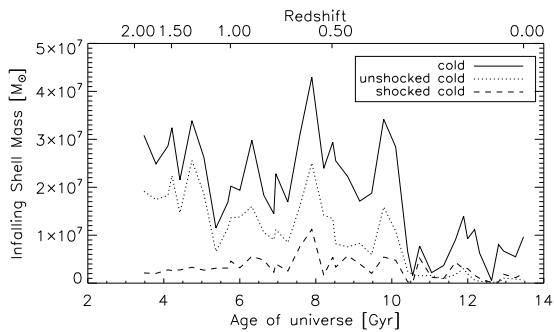}
\caption{Mass of infalling gas with $T < 10^5$ K found in a 1 kpc-wide shell at 30 kpc from the centre. Assuming that this gas will reach the disc in a short time, we may use it as a proxy for an accretion rate onto the disc. The sharp drop at $\sim11$ Gyr is coincident with the disappearance of the warp, therefore implying that the appearance of a large-scale warp requires ongoing accretion of cold gas.}
\label{fig:disk_accretion}
\end{figure}

To test this hypothesis, we evaluate how much gas is being supplied to the disc as a function of time. In Figure~\ref{fig:disk_accretion} we show the mass of cold gas falling toward the centre through a 1 kpc shell at 30 kpc (comoving) away from the disc. The shell is chosen to be far enough from the centre that it does not include any part of the disc, yet close enough that we can reasonably expect the cold gas in this shell to make it to the disc in a short time. We show the contribution of unshocked and shocked particles to this cold gas reservoir by dotted and dashed lines respectively; the remainder is made up of gas that enters the halo as a part of a merging event. 

%
%

Figure~\ref{fig:disk_accretion} shows that a significant drop in the amount of disc fuel coincides with the disappearance of the warp, confirming our expectations that the warp is kept alive by continual accretion. Unshocked gas is most important for disc growth in this galaxy (see Figure 7 of \citealt{Brooks:2009}), therefore it follows that the decrease in the unshocked fraction has the largest effect on the warp.  Further contributing to the disappearance of the warp is the growth of the central disc. As it grows, it begins to dominate regions where the warp had been prominent before.

We note that satellite accretion events or perturbations do not necessarily result in large-scale warps. At $\sim12$ Gyr, there is an influx of gas stripped from a satellite, shown by the excess cold gas which is not attributed to either shocked or unshocked components, but the perturbation fails to incite a warp. The cold gas contributed by the satellite by itself does not reach requisite densities to appear in the warp.   

%
%

\section{Properties of Stars in the Misaligned Disc}
\label{sec:misalignedstars}

Extended star formation in the misaligned gas component must leave a trace in the stellar populations at the
present day. We can begin to investigate the importance of stars forming in the misaligned component by 
dividing stars based on whether they were formed interior or exterior to the warp radius at their time of formation. Determining the galactocentric radius of formation in the disc plane is not trivial in a cosmological simulation because neither the disc's centre nor its orientation in the simulation frame are fixed. Further, the variations of the disc's orientation are temporally poorly sampled because for practical reasons we only retain a limited number of outputs. Simply 
extrapolating the disc's position and orientation at each snapshot and the time of formation of stars proves insufficient. 

However, we do retain full information about the positions and times of star formation events for the entire simulation, regardless of when the outputs are produced. To proceed, we make the assumption that in the main galaxy, the vast majority of the star formation occurs in the disc and is symmetric about the disc's centre. With this assumption, and given a sufficient amount of star formation, we now have knowledge of the disc's 3D orientation within the simulation volume, sampled at the star formation timescale ($\sim1$ Myr in our recipe). We bin stars according to their time of formation in 10 Myr bins and for each bin determine the disc plane. With this finely sampled orientation information, we then determine the precise radius of formation for each star in the plane of the disc at that time.

\begin{figure}
\centering
\includegraphics[width=3in]{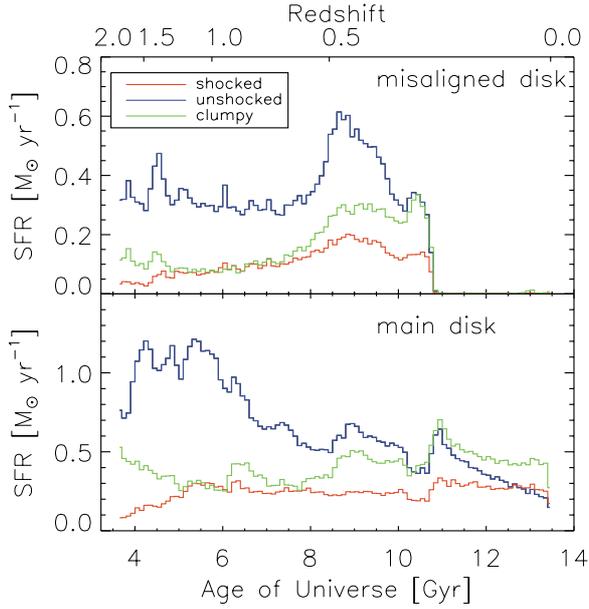}
\caption{Star formation rate as a function of time in the misaligned and main star-forming disc. In blue
we show the star formation rate from cold gas accretion, in red from hot accretion, and in green from 
clumpy accretion. See text for definitions.}
\label{fig:sfrcomps}
\end{figure}

In Figure~\ref{fig:sfrcomps} we show the star formation history in the misaligned disc and main star-forming disc. We restrict our analysis to stars with disc-like kinematics at z=0, defined to be stars that satisfy $J_z/J_{circ}(E) > 0.8$, where $J_{circ}(E)$ is the angular momentum of a circular orbit at the particle's energy.  Colours correspond to the accretion mode of gas, as in Figure~\ref{fig:discmass}.

Figure~\ref{fig:sfrcomps} shows that the unshocked gas is the dominant fuel for star formation during most of the evolution, in both the main and misaligned discs, which disappears almost completely at 11 Gyr. Until the last few Gyr, the SFR in the misaligned disc is lower than in the main disc by factor $\sim 2$, although the total gas mass is higher (see Figure~\ref{fig:discmass}). This is due to overall low gas densities in the outer misaligned component and correspondingly low star formation efficiency.  

\begin{figure}
\centering
\includegraphics[width=3in]{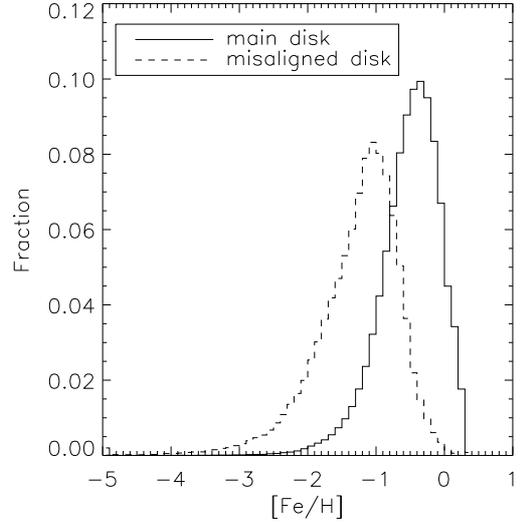}
\caption{Mass-weighted metallicity distribution function for the stars found in the main disc (solid line) and the misaligned disc (dashed line).}
\label{fig:metaldist}
\end{figure}

The inefficient SFR in the misaligned disc implies that the enrichment of the gas in the misaligned disc will proceed much more slowly than in the main disc, resulting in a more metal-poor population. In addition, whatever enrichment takes place, the metallicity of the misaligned disc is continually diluted by the infall of pristine gas, further keeping the metallicity low. We show this difference explicitly in Figure~\ref{fig:metaldist}, where we plot the metallicity distribution function (MDF) for the stars forming in the main disc and those in the misaligned disc. The offset in the peaks of the distributions is $\sim0.8$ dex. The detailed differences between the two MDFs will vary depending on the properties of the misaligned disc. However, the short life-cycle of gas in the misaligned disc should be general and will therefore always result in low enrichment. 

%
%

\begin{figure}
\centering
\includegraphics[width=3in]{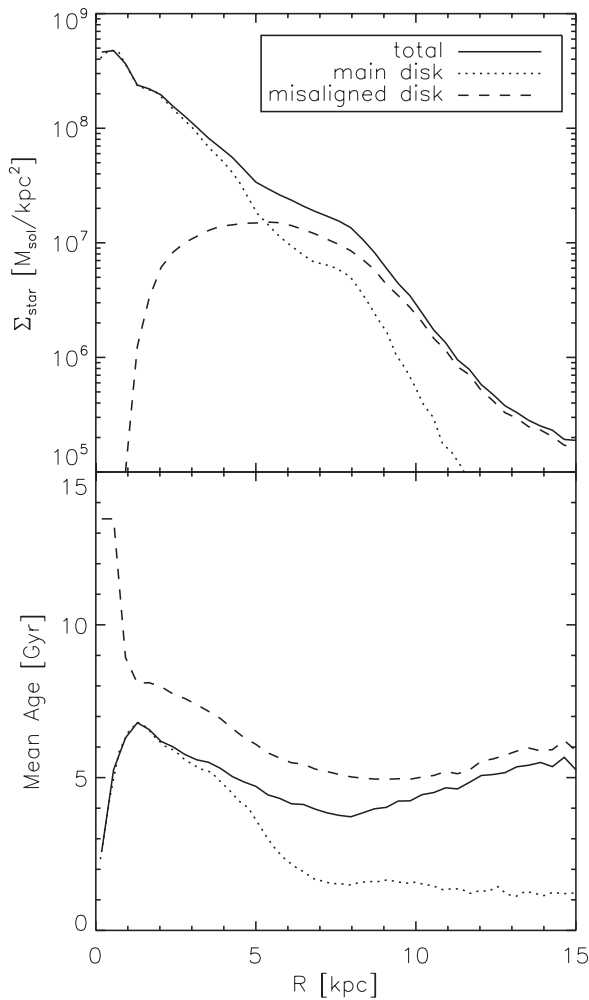}
\caption{{\bf Top:} Surface density profile showing all disc stars (solid), stars formed in the misaligned component 
(dashed) and stars formed in the plane of the main disc (dotted). Stars formed in the warp dominate the outer parts of the disc. The absence of stars formed out of the plane at low radii is due to our method of selection.
{\bf Bottom:} Mean age profile of disc stars.}
\label{fig:profiles}
\end{figure}

The top panel of Figure~\ref{fig:profiles} shows the z=0 radial surface density profiles of all of the star particles 
making up the disc (solid line), decomposed into those stars that formed in the misaligned disc (dashed) and 
those formed in the main star-forming disc (dotted). The absence of misaligned-disc stars in the centre is due to 
our definition of misaligned disc stars as having $R_{birth} > 3$ kpc. Stars that formed in the misaligned disc completely dominate the present-day outer disc. This is natural if we consider that the misaligned disc component persists for 7 Gyr and is by definition always beyond the main star-forming disc. 

%
%

Figure~\ref{fig:profiles} suggests that the stars forming in the warp dominate the present-day surface density at $R \ga 6$ kpc, or roughly 2 scalelengths. This is partly due to the fact that during the lifespan of the warp, the main disc region extends only to $\sim5$ kpc (only $\sim2$~scale lengths). The warp may therefore be dominant in a larger fraction of the disc in this model than in some local examples, where warps are seen to begin only at several disc scale lengths. Consequently, the amount of stars formed in the warp is likely  too high in this model, and their contribution to the thick and extended stellar population may be unusually large. However, we do not have constraints from more distant systems, where it is possible that warps begin at smaller galactocentric radii.  

Stars formed in the warp dominate the stellar density at larger radii, and because their star formation history is rather extended, the mean age profile shows a slight upturn in the outer disc. This upturn in the age profile is similar to that in R08, which was caused instead by radial migration. In cosmological AMR simulations, \citet{Sanchez-Blazquez:2009} also found that a change in the age profile was due to prolonged star formation in the outer disc. However, as we discuss in $\S$~\ref{sec:dynamics} below, we cannot use the present models to determine the relative importance of stars forming in the warp to stars that may have migrated there, because the discs presented here do not support the spiral structure required for migration. For comparison, the migrated population at 15 kpc in R08 resulted in a surface density of $\sim10^6 \text{M}_{\odot}\text{ kpc}^{-2}$, which is an order of magnitude higher than the amount of stars we find forming in the misaligned disc. We therefore stress that while stars may indeed form in the outer disc over extended periods of time, given a disc that produces realistic dynamical behavior, the migrated population should contribute substantially to the stellar density in the outermost parts of the disc. However, in lower mass spirals where we may expect spiral structure to be less important than in Milky Way-size discs, the migrated population may well be sub-dominant in the outer disc. 

The star formation in the warp is also problematic because much of it occurs near the gas density limit imposed by the sub-grid star-formation prescription. We must therefore allow for the possibility that star formation in the warp is somewhat overestimated in our model. However, large-scale warps such as the one described in this Paper are common in many cosmological simulations. Because of their potential to dominate over the in-plane population at a few scale-lengths, stars formed in the warp should be treated with caution, especially when analyzing the overall kinematic properties of discs formed in cosmological simulations as they are likely to exacerbate the problematic high velocity dispersions (see $\S$~\ref{sec:dynamics}).

\section{Thick discs from Outer disc Misalignments}
\label{sec:thickdisk}

\begin{figure*}
\centering
\includegraphics[width=6.5in]{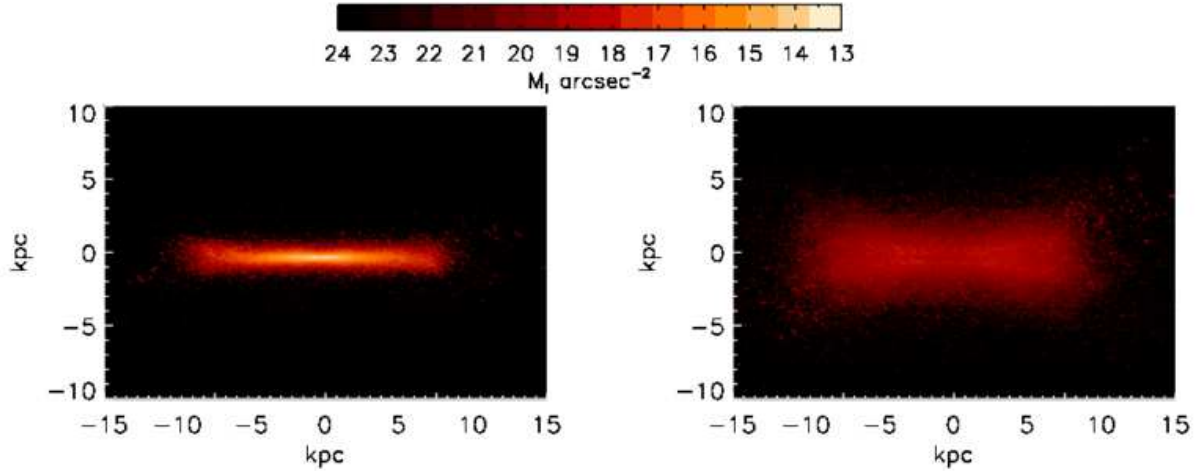}
\caption{Surface brightness maps of disc stars at $z$=0. The left panel shows stars formed in the main star 
forming disc, while the right panel shows stars formed in the misaligned component. Although the stars on the 
right were formed in-situ and have disc-like kinematics, due to their extraplanar origin they result in a thicker,  
lower surface brightness present-day population.}
\label{fig:warp_maps}
\end{figure*}

The dominance of stars that formed from the misaligned gas component in the outer parts of the disc in our simulation opens the interesting possibility that such stars could contribute substantially to the present-day out-of-plane population. Because the disc grows with time, what was the outer disc at z=1 is a modest distance from the centre at z=0. In Figure~\ref{fig:warp_maps} we show SUNRISE \citep{Jonsson:2006} edge-on I-band images of disc stars at z=0, split in the same way as in Figure~\ref{fig:profiles}. The left and right panels show stars formed in the main star-forming disc and in the misaligned component respectively. Although they have disc-like kinematics, the stars which formed out of the disc plane constitute a much thicker present-day population than the stars born in the main disc. Though stars forming in the warp at any given time are in a tilted disc, their orbits precess resulting in a thickened population by present day rather than an inclined disc. We defer making detailed comparisons with observed thick discs to future work, but we point out that they could be significantly affected by long-lived warps. 

Given the tenuous nature of the misaligned component manifested in its low star formation efficiency, the amount of stars forming out of that gas is somewhat dependent on our choice of star formation parameters. The parameter set used here has been shown previously to yield excellent agreement with observations in terms of the Schmidt-Kennicutt law \citep{Stinson:2006}, mass-metallicity relation \citep{Brooks:2007} and the Tully-Fisher relation \citep{Governato:2009}, providing evidence that the SF recipe used with these parameters is remarkably successful in a globally-averaged sense. However, star formation in such low-density regions in the observed universe may occur in localized clumps (e.g. \citealt{Ferguson:1998, Thilker:2005, Dong:2008}) and it is not clear that at the present state of simulations we can satisfactorily resolve such processes for Milky Way-size galaxies. Recent attempts to resolve molecular cloud masses in cosmological simulations yielded promising results \citep{Agertz:2009, Ceverino:2009, Governato:2010}, but due to computational cost only the simulations of dwarfs have been carried out to z = 0. To fully understand the impact of the misaligned component on the disc stellar populations, a full parameter study of the star formation prescription is required, using higher resolution simulations than those currently available to us, where star formation will be resolved in localized clumps.

\section{Present-day Observable Consequences of Misaligned Accretion}
\label{sec:present_day_obs}

In the previous sections we discussed the consequences of star formation in the warp for a system whose warp disappeared a few Gyr ago. However, many galaxies in the local universe show large-scale gas warps, so we now turn to a simulation whose warp is preserved to the present-day. The details of this model are described in $\S$~\ref{sec:methods}. We have verified that the physical reasons for the warp in this model are analogous to ones in our fiducial simulation described in the preceding sections. Although we do not show it here, a Briggs figure for this model (like those shown in Fig.~\ref{fig:briggs}) traces out a smooth leading spiral. Below, we describe the observational predictions manifesting in a unique stellar population signature from warps forming by misaligned gas infall. These predictions should make it possible to directly determine whether nearby large-scale warps are caused by such a mechanism and therefore confirm the presence of both, current cold accretion and the existence of a hot gas halo.

\begin{figure*}
\centering
\includegraphics[width=6.5in]{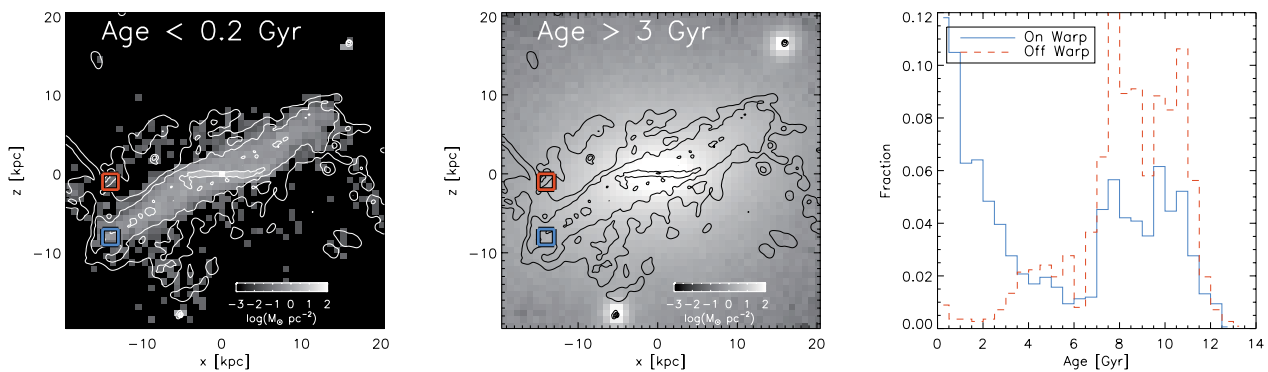}
\caption{Projected stellar surface density distribution for young (left panel) and old (middle panel) stellar populations. Contours show neutral hydrogen column density. The right panel shows the age distribution within the blue (solid, on-warp) and red (dashed, off-warp) boxes respectively. An old halo population exists in both, but the on-warp stellar populations are dominated by recent star formation. 
}
\label{fig:age_map}
\end{figure*}

In Figure~\ref{fig:age_map} we show the projected stellar density of stars in this model at present day with overlaid HI column density contours. We split the stars into young (age $< 200$ Myr) and old (age $> 3$ Gyr) and show the corresponding maps in left and middle panels respectively. Resolved star studies using the Hubble Space Telescope (HST) can discern between such stellar populations using CMD modeling (e.g. \citealt{de-jong:2007}). The young stars are predominantly found in the main disc and the HI warp, while the old stars dominate the more extended and largely spherical halo. 

In the right panel of Fig.~\ref{fig:age_map}, we show the age distributions of the two small boxes shown in the left and middle panels. The age distribution of the blue box, coincident with the warp, is dominated by recent star formation, though an older halo population is also present. The tilted and extended old population can be seen in the vicinity of the current warp in the middle panel, likely a relic of star formation that had taken place in the misaligned component in the past. The off-warp box is completely dominated by an old population. 

Evidence for the processes described in this Paper may therefore be found in similar large-scale warps, which have recently been associated with low levels of star formation in the outermost disc regions of M83 and NGC5055  \citep{Thilker:2005, Sancisi:2008}. Spectroscopic studies of ionized gas in the XUV disc of M83 indicate that the warped gas has low metallicity (0.1 - 0.2 $Z_{\odot}$) consistent with enrichment from a stellar population younger than 1 Gyr \citep{Gil-de-Paz:2007}. A similar result has recently been reported in a spectroscopic study of a polar disc galaxy NGC4650A \citep{Spavone:2010}. Studies of resolved stellar populations in and surrounding nearby warps could distinguish between different warp formation mechanisms and provide insight about the importance of on-going cold accretion and hot gas halos in massive spirals, both of which are inherently very difficult to observe directly.

\section{Disc Heating and Dynamics}
\label{sec:dynamics}

It is important to understand the relative importance of stars forming in the misaligned disc compared to other mechanisms that may also be responsible for substantial amounts of material in the outer disc. In particular, R08 showed that radial migration in a growing galactic disc can efficiently populate the outer disc, making the outer disc a potentially valuable laboratory for studying such processes. Can we use the cosmological simulations to understand how secular evolution compares to the processes operating on a larger scale? Radial migration results from the interaction of star particles with transient spiral arms, formed through gravitational instabilities \citep{Sellwood:2002}. However, in the absence of a cooling mechanism,  a single spiral will heat the disc enough to stabilize it completely within a few rotation periods \citep{Sellwood:1984}. In the simulations of R08, the disc was cooled by the continual accretion of gas and formation of kinematically cold stars, thus remaining at the cusp of instability and providing conditions appropriate for continual transient spiral triggering. R08 showed that even when we consider the change in radius for particles on mostly circular orbits we see that many particles are far from their birth radii by the end of the simulation. Thus the observed differences between the galactocentric radii of stars at the end of the simulation and their birth locations are not only because of orbital heating, but rather due to significant changes in the guiding centers of the stars. 

\begin{figure}
\centering
\includegraphics[width=3in]{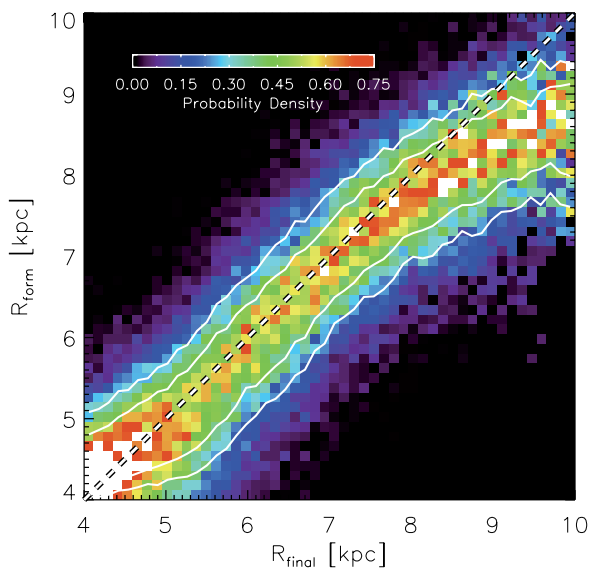}
\caption{
Probability density of formation radius as a function of final radius for stars that formed in the main disc plane. 
Cell values are scaled such that an integral over the values in each column multiplied by the area of a cell equals unity. The white contours enclose 50 per cent and 75 per cent of the mass at each given 
$R_{final}$. We find little evidence for significant radial redistribution of stars.
}
\label{fig:rfinalrform}
\end{figure}

We now try to determine whether the disc simulated in a fully cosmological setting described in the preceding sections is subject to the same radial migration mechanisms seen in the idealized simulations of R08. Figure~\ref{fig:rfinalrform} shows the conditional probability distribution of formation radii given the final radii for particles formed in the main star-forming disc. We only plot a kinematically-cool subset of the disc particles, defined by having $J_z/J_{circ}(E) > 0.95$, where $J_{circ}(E)$ is the maximum angular momentum in the disc at that particle's energy. This selection ensures that the spread around the 1:1 line in Figure~\ref{fig:rfinalrform} is due to secular redistribution of star guiding centers rather than disc heating. Figure~\ref{fig:rfinalrform} shows that very little radial migration has taken place in the presented model, with the 75 per cent mass contour indicating at most 1 kpc difference between $R_{final}$ and $R_{form}$ (in the simulations of R08, this difference is $\sim4$~kpc at 8 kpc). This differs from the recently reported results studying radial migration in AMR simulations \citep{Sanchez-Blazquez:2009, Martinez-Serrano:2009}, although we emphasize that care must be taken in cosmological simulations to distinguish heating from migration. 

%
%

\begin{figure}
\centering
\includegraphics[width=3in]{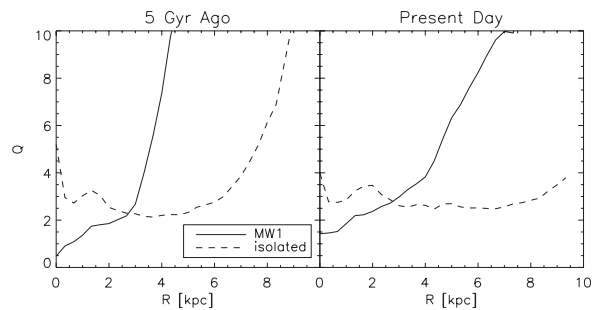}
\caption{Toomre Q parameter as a function of radius for the MW1 disc at two different epochs, compared 
with the Q measured in the isolated disc of R08. The cosmological disc is kinematically much hotter 
at most radii, which prevents the formation of spiral structure.}
\label{fig:toomreq}
\end{figure}

Given that the simulations of R08 were idealized models, it is important to examine the underlying reasons for the different behavior found in the cosmological simulation. It is a known problem of cosmological simulations that the resulting discs are too kinematically hot, suppressing the formation of asymmetries. In Figure~\ref{fig:toomreq} we show the Toomre Q stability parameter \citep{Binney:2008}
\begin{equation}
Q \equiv \frac{\sigma_{R} \kappa}{3.36 \Sigma G}
\end{equation}
for the disc of R08 and the MW1 disc 5 Gyr ago and at present-day. We calculate Q using only the stellar component. The Q parameter for the isolated disc is between 2-3 (marginally stable). \citet{Sellwood:1984} found that in a model where the disc is continually cooled, the value of Q reached a quasi steady-state of $\ga 2$, which allowed for continual formation of transient spirals, in qualitative agreement with the behavior seen in the R08 isolated simulation. The disc of MW1, however, heats to significantly higher values at modest radii and most of the disc is at Q $> 3$ even at present day when the disc extends to $\ga 10$ kpc. Such a disc cannot be expected to support spiral structure required for radial migration. Indeed, the amplitude of the m=2 mode for this disc is $< 0.02$ everywhere beyond the bar at z=0, a degree of azimuthal asymmetry appropriate for early-type discs \citep{Rix:1995}. In the model of R08, the m=2 amplitude ranges from 0.02 to 0.1 at the end of the simulation. The total velocity dispersion ($\sigma_{tot}^2 = \sigma_r^2 + \sigma_{\phi}^2 + \sigma_{z}^2$) for a 1 kpc ring centered at 8 kpc exceeds 100 km~s$^{-1}$ for the cosmological model, which is much higher than the measured value of 70 km~s$^{-1}$ for the oldest stars in the solar neighborhood \citep{Holmberg:2009}.

%
%

The current state of cosmological simulations available to us therefore does not allow us to address the relative importance of radial migration compared to star formation in warps for the outer disc stellar populations. However, we feel justified in addressing the more general issues of angular momentum alignment and accretion, since the relevant processes are seeded in the larger-scale cosmological dynamics. 

\section{Discussion and Conclusions}
\label{sec:discussion}

The high frequency of warps in the nearby universe (e.g. \citealt{Garcia-Ruiz:2002}) warrants efforts 
to understand the phenomenon in some detail. The range of warp morphologies is vast,  though empirical rules do exist for the behavior of observed warps \citep{Briggs:1990}. In this Paper, we presented a possible formation mechanism for large-scale warping of disc gas in a Milky-Way size galaxy, similar in spirit to previous theories in that it relies on cosmological accretion of material. However, virtually all previous theories regarding warps postulate the warping of disc material \emph{after} it had settled in the main disc plane (e.g. \citealt{Toomre:1983, Ostriker:1989, Debattista:1999, Shen:2006, Weinberg:2006, Dubinski:2009}). In this Paper, we have described a fundamentally different warping mechanism, one which ties directly the morphological appearance of the gas disc to the angular momentum properties of the (invisible) gas halo. In this mechanism, the warp is a result of the hot gas halo torquing the infalling cold gas into a misalignment with the central disc. The central disk and the hot gas halo are misaligned because they respond differently to accretion and substructure perturbations. The warp is therefore at once an indication of ongoing cold gas accretion as well as the spin of the hot gas halo outside the central region. 

``Cold accretion'' discussed here should not be confused with high-redshift ``cold streams'' (e.g. \citealt{Dekel:2009, Agertz:2009, Ceverino:2009}), which are a result of filaments penetrating the virialized halo directly to the central disc region. The cold accretion at low redshift is instead in cold clouds and not filamentary in nature inside the virial radius. Nevertheless, the fact that some gas enters the halo at relatively low temperatures $\la 10^5$K has large consequences for the disc-building process, as evidenced in the preceding sections. 

Not all galaxies show warps of quite the same magnitude as the one we have described in this Paper, and 
in particular the current amplitude of the Milky Way warp appears smaller. We restricted our discussion
to the large warping specifically to focus on its cosmological implications. Mechanisms for
generating smaller vertical offsets in the disc plane appropriate for the current state of the Milky Way warp 
have been discussed by many other authors (e.g. \citealt{Toomre:1983, Ostriker:1989, Debattista:1999, 
Shen:2006, Weinberg:2006, Dubinski:2009}). 

Although our own Galaxy does not show such a prominent warp, extragalactic systems of this sort abound 
in the nearby universe. Much effort has in recent years been devoted to uncovering sources of disc fueling, 
such as isolated high-velocity clouds (HVCs) either around the Milky Way or other galaxies. The general 
conclusion is that observed sources of HI gas do not account for sufficient accretion rates to sustain current observed SFRs \citep{Sancisi:2008}. We have shown that the large warp requires a constant supply of fresh cool gas (Figure~\ref{fig:disk_accretion}), and argued that the lifespan of individual parcels of gas in the outer disc is short. Thus, the presence of such a feature should be considered together with other indicators of accretion, such as discrete cold clouds, as evidence for ongoing cosmological gas infall. 

We have shown in $\S$~\ref{sec:present_day_obs} that the warp formation process described in this Paper may be observable in nearby systems. Specifically, we have argued that the connection between XUV discs and HI warps is a direct result of the warping processes described in this Paper. Furthermore, the unique stellar population signature imprinted by this particular warping process should be discernible in inclined nearby systems where resolved star observations of outer discs are possible. 

We stress, that large-scale warps and their effects on stellar populations have largely gone unnoticed in recent work on cosmological disc galaxy formation, although warps readily form in most such models. The amount of stars forming in the misaligned disc is likely model-dependent regardless of the simulation method used, but stars formed in this way can drastically alter the stellar population properties in the outer parts of the disc, including kinematics and metallicity (see $\S$~\ref{sec:misalignedstars} and $\S$~\ref{sec:dynamics}). Although we used a single simulation in this work to describe the physical process responsible for the warp, many cosmological simulations of isolated spiral galaxy formation yield a large-scale warp during some part of the galaxy's evolution. The details of the misalignment (longevity, degree of tilt) will differ due to different merger histories, but the general features outlined here should be valid for all systems of similar mass.

\section*{Acknowledgments}
The authors thank the hospitality of KITP at UCSB where this work was initiated. R. R. thanks A. Stilp for the use of her code to produce HI maps and P. Jonsson for his SUNRISE radiative transfer code and comments on the manuscript. R. R. also thanks R. de Jong for fruitful discussions regarding warps. R. R. was supported by the NSF grant CAREER AST 02-38683 (also partially supporting J. J. D.) and NSF ITR grant PHY 02-05413 (also partially supporting T. R. Q. and F. G.) at the University of Washington. V. P. D. was supported by an RCUK Fellowship at the University of Central Lancashire. A. B. acknowledges funding from the Sherman Fairchild
Foundation. F. G. acknowledges support from HST GO-1125, NSF grant AST-0607819 and NASA ATP NNX08AG84G. Simulations were carried out on TeraGrid resources at TACC, SDSC, and PSC as well as additional resources at ARSC. Analysis was performed on resources purchased with the help of the University of Washington Student Technology Fee. 

\bibliographystyle{mn2e}

\end{document}